\documentclass[letterpaper]{article} 
\usepackage{aaai23}  
\usepackage{times}  
\usepackage{helvet}  
\usepackage{courier}  
\usepackage[hyphens]{url}  
\usepackage{graphicx} 
\usepackage{natbib}  
\usepackage{caption} 
\usepackage{algorithm}

\usepackage{subfigure}
\usepackage{amsmath}
\usepackage{multirow,booktabs,setspace,caption}
\usepackage{algpseudocode}
\usepackage[usenames,dvipsnames]{color}
\usepackage{varwidth}

\usepackage{newfloat}
\usepackage{listings}
\DeclareCaptionStyle{ruled}{labelfont=normalfont,labelsep=colon,strut=off} 
\floatstyle{ruled}
\newfloat{listing}{tb}{lst}{}
\floatname{listing}{Listing}
\title{Semantic Interactive Learning for Text Classification: A Constructive Approach for Contextual Interactions}
\author{
    Sebastian Kiefer,
    Mareike Hoffmann
}
\affiliations{
    \textsuperscript{}Cognitive Systems, University of Bamberg, Germany\\

}

\usepackage{bibentry}

\begin{document}
\nocopyright
\frenchspacing
\maketitle

\begin{abstract}
Interactive Machine Learning (IML) shall enable intelligent systems to interactively learn from their end-users, and is quickly becoming more and more important. Although it puts the human in the loop, interactions are mostly performed via mutual explanations that miss contextual information. Furthermore, current model-agnostic IML strategies like CAIPI are limited to 'destructive' feedback, meaning they solely allow an expert to prevent a learner from using irrelevant features. In this work, we propose a novel interaction framework called \textit{Semantic Interactive Learning} for the text domain. We frame the problem of incorporating constructive and contextual feedback into the learner as a task to find an architecture that (a) enables more semantic alignment between humans and machines and (b) at the same time helps to maintain statistical characteristics of the input domain when generating user-defined counterexamples based on meaningful corrections. Therefore, we introduce a technique called \mbox{SemanticPush} that is effective for translating conceptual corrections of humans to non-extrapolating training examples such that the learner's reasoning is pushed towards the desired behavior. In several experiments, we show that our method clearly outperforms CAIPI, a state of the art IML strategy, in terms of Predictive Performance as well as Local Explanation Quality in downstream multi-class classification tasks.

\end{abstract}

\section{Introduction} \label{Introduction}
Although modern ML approaches improved tremendously with regard to prediction accuracy and even exceed human performance in many tasks, they often lack the ability to allow humans develop an understanding of the whole logic or of the model's specific behavior \citep{Adadi2018, Holzinger2016, Holzinger2017}. Additionally, most systems don't allow the integration of corrective feedback used for model adaptation.
Consequently, different research disciplines have emerged that provide first solutions. \textit{Interpretable Machine Learning} as well as \textit{Explainable Artificial Intelligence}, which can be summarized as \textit{Comprehensible Artificial Intelligence} \citep{Bruckert2020} when being combined, shall allow for global or local interpretability as well as transparent and comprehensible ML results \citep{Akata2020}.
Nevertheless, the according explanations used for better transparency and human comprehensibility during human-machine-interactions are mostly considered unidirectional, from the AI system to the human and often lacking contextual information \citep{Adadi2018}.
Therefore, any correction of erroneous behavior or any inclusion of domain-specific knowledge through human experts is not possible in a model-agnostic way \citep{Bruckert2020}.
\textit{Explanatory Interactive Machine Learning} addresses that with the intention to 'close the loop' by allowing human feedback in the form of machine-to-user explanations based on transparent decisions \citep{Teso2019}.
The authors from \citep{Teso2019} demonstrated that not only the predictive and explanatory performance of a learner, but also the process of building trust in a learner benefit from interactions through explanations.
Except from systems like EluciDebug \citep{Kulesza2015} or Crayon \citep{Fails2003}, which use feedback to adapt a learner (albeit model-specific), there are still few possibilities for holistic, meaningful and model-agnostic interventions to correct learners' mistakes and use expert knowledge.
Based on this research gap, we phrase the following research questions (RQ):
(1) How can we develop a model-agnostic Interactive ML approach that offers semantic (constructive, meaningful, contextual, and realistic) means for performing corrections and providing hints?
(2) Will the elaborated interactive system converge faster to higher learning performance with the help of contextual explanations and the new interaction paradigm than state-of-the-art methods?
(3) Will the generated explanations of our method quicker be conclusive than those from state-of-the-art methods?
During our research, we propose an architecture called Semantic Interactive Learning and evaluate its quality with regard to these research questions.


%
%

\section{Related Work} \label{Related Work}
\textit{Human-Centered Machine Learning} can be summarized as methods of aligning machine learning systems with human goals, context, concerns, and ways of work \citep{Gillies2016}. It is strongly connected with Interactive Machine Learning as an interaction paradigm in which a user or user group iteratively trains a model by selecting, labeling, and/or generating training examples to deliver a desired function \citep{Dudley2018}. One can assume that a learner could better be aligned with human goals if the end user knows more about its behavior (Explanatory Interactive Machine Learning). \citeauthor{Kulesza2015} (\citeyear{Kulesza2015}) proved that intuition with their Explanatory Debugging approach. They additionally showed that not only the machine benefits from corrections based on transparent explanations, but also the user was able to build a more accurate mental model about the learners behavior. 

Hence, interactions between humans and machines via mutual explanations \citep{Schmid2020, Bruckert2020} have the potential to adequately bring the human in the loop in a model-agnostic way.
The overall process should work as a Training-Feedback-Correction-cycle that enables a Machine Learning model to quickly focus onto a desired behavior \citep{Fails2003}. Users should be able to iteratively integrate corrective feedback into a Machine Learning model after having analyzed its decisions \citep{Amershi2014}.

Consequently, \citeauthor{Teso2019} (\citeyear{Teso2019}) included a local explainer called \textit{Local Interpretable Model-Agnostic Explanations} (LIME) into an active learning setting. Their framework proposes a method called CAIPI which enables users to correct a learner when its predictions are right for the wrong reasons by adding counterexamples in a 'destructive' manner. The correction approach is based on \citeauthor{Zaidan2007} (\citeyear{Zaidan2007}). As an example from the text domain, words which are falsely identified as relevant get masked from the original document and the resulting counterexamples recur as additional training documents.

Although CAIPI paves a first way for model-agnostic and explanatory IML, it reveals some significant drawbacks.
First, it only operates by deleting irrelevant explanations (What has incorrectly been learned?). Thus, it is limited to a 'destructive' feedback about incorrectly learned correlations while an active learning setting might rarely contain right predictions made for the wrong reasons. 
Second, CAIPI uses contextless explanations as a basis and in turn applies contextless feedback by independently removing irrelevant explanatory features. Doing so, human conceptual knowledge might hardly be considered during interactions, although it is known, that harnessing conceptual knowledge “as a guiding model of reality” might help to develop more explainable and robust ML models, which are less biased \citep{Holzinger2021}. A first step towards that was suggested by \citeauthor{Kiefer2022} (\citeyear{Kiefer2022}). In the according research, topicLIME is proposed as an extension of LIME that offers contextual and locally faithful explanations by considering higher-level semantic characteristics of the input domain within the local surrogate explanation models.
As third drawback, CAIPI enables 'discrete' feedback only. In the domain of text, it is based on mutual explanations in bag-of-words representation, where words as explanatory features are either present or not. Therefore, graded, continuous feedback is not possible.

When explaining and correcting a classifier in a way as described above, neighborhood extrapolation to feature areas with low data density, especially in case of dependent features \citep{Molnar2019}, causes a classifier to train on contextless counterexamples sampled from unrealistic local perturbation distributions. This circumstance might lead to generalization errors.

The overall goal of this work is therefore to enable more realistic and constructive interactions via semantic alignment between humans and ML models across all possible types of a learner's reasoning and prediction errors.

%
%
%
%
%
%
%

\section{Method}
\begin{figure}[ht!]
	\includegraphics[width=0.475\textwidth]{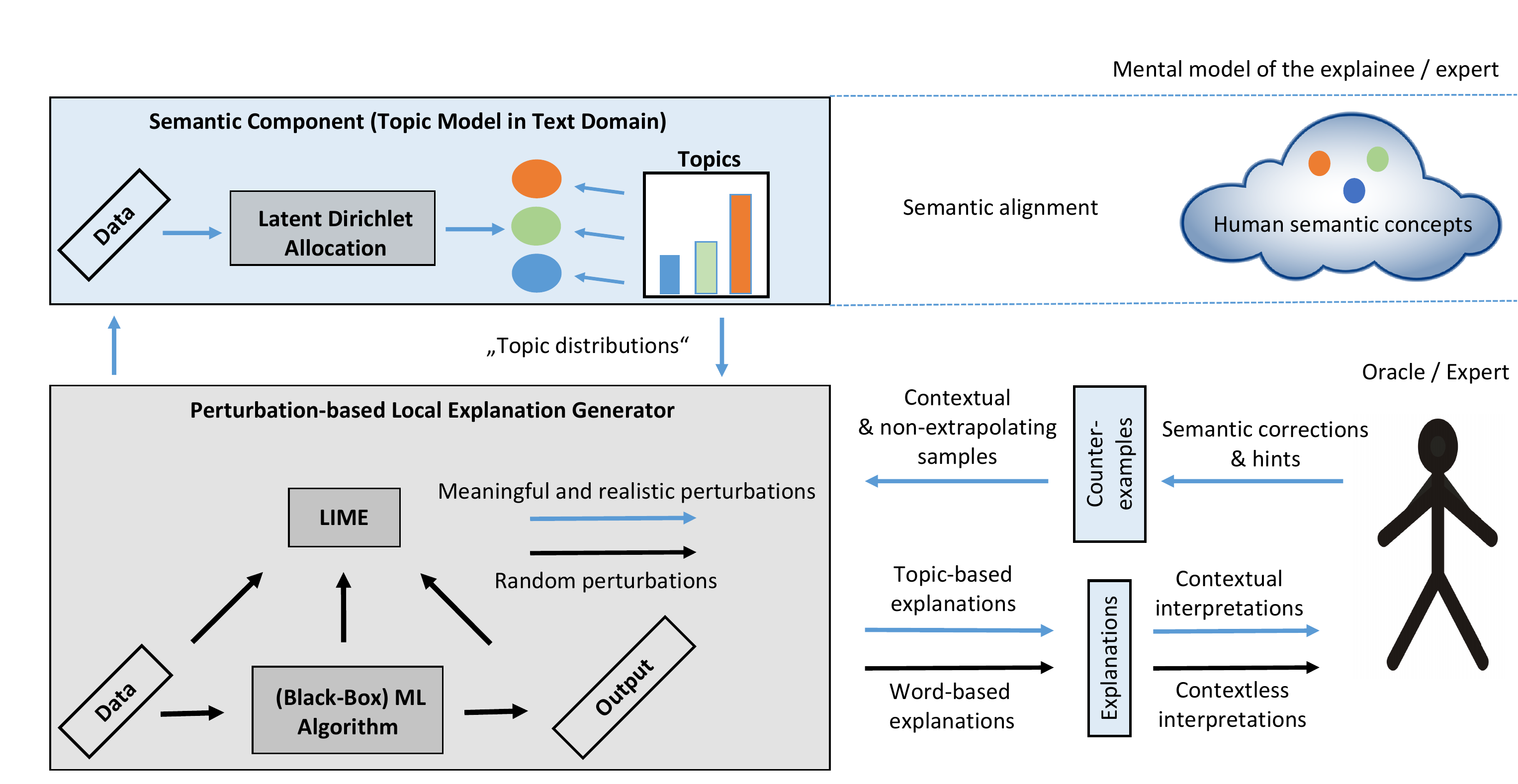}
	\caption{Architecture for constructive and contextual interactions.}
	\label{img:CaSEArchitecture}
\end{figure}

Figure \ref{img:CaSEArchitecture} depicts our proposal for answering RQ1 from architectural point of view.
It extends previous research called \textit{Contextual and Semantic Explanations} (CaSE) \citep{Kiefer2022}. CaSE suggests a framework that allows humans contextual interpretations of ML decisions in a model-agnostic way via topic-based explanations. While CaSE solely refers to the process of explanation generation, our research aims at closing the loop and enabling humans to integrate domain knowledge via semantic corrections and hints. The following subsections briefly describe the components contained in our framework and especially introduce our new IML strategy called SemanticPush.

\subsection{Latent Dirichlet Allocation} \label{LDA}
We instantiate the semantic component of our framework with a method called Latent Dirichlet Allocation (LDA).
It can be described as a hierarchical Bayesian model for collections of discrete data \citep{Blei2003}.
Used in text modeling, it finds short representations of a corpus' documents and preserves essential statistical relationships necessary for making sense of the input data. After training, each document can be characterized as a multinomial distribution over so-called topics.
For each document \textbf{w} in a corpus \textbf{D}, a generative process, from which the according documents have been created, is assumed:

\begin{enumerate} \label{enum_LDA}
\scriptsize

\item Choose N (the number of words) $\sim$ Poisson($\xi$).
\item Choose $\theta$ (a topic mixture) $\sim$  Dir($\alpha$).
\item For each of the N words $w$\textsubscript{n}:
\begin{enumerate}
\item Choose a topic $z$\textsubscript{n} $\sim$ Multinomial($\theta$).
\item Choose a word $w$\textsubscript{n} from $p($w$\textsubscript{n} \vert z\textsubscript{n}, \beta)$, a multinomial probability conditioned on the topic $z\textsubscript{n}$.
\end{enumerate}
\end{enumerate}

The joint distribution of a topic mixture $\theta$, a set of topics \textbf{z}, and a set of words \textbf{w} given the hyperparameters $\alpha$ and $\beta$ is characterized by:

\begin{equation} \label{eq:1}
\scriptsize
p(\theta,\textbf{z},\textbf{w} \vert \alpha, \beta) = p(\theta \vert \alpha) \prod \limits_{n=1}^{N}p(z\textsubscript{n}\vert \theta)p(w\textsubscript{n}\vert z\textsubscript{n},\beta).
\end{equation}

We combine LDA with a coherence measure called C\textsubscript{v} coherence. It is used for finding an appropriate hyperparameter \textit{number of topics k} that LDA shall infer.
Röder et al. found that coherence measure to be the best in terms of its correlation with respect to human topic-interpretability \citep{Roeder2015, Syed2017}.

\subsection{LIME and topicLIME}
Ribeiro et al. developed LIME, a method that explains a prediction by locally approximating the classifier's decision boundary in the neighborhbood of the given instance \citep{Ribeiro2016a}.
The final objective is to minimize a measure $\mathcal{L}(f, g, \pi_x(z))$ that evaluates how unfaithful $g$ (the local explanation model) is in approximating $f$ (the model to be explained) in the locality defined by $\pi_x(z)$. Striving for both interpretability and local fidelity, an explanation is obtained by minimizing $\mathcal{L}(f, g, \pi_x(z))$ as well as keeping $\Omega(g)$ low enough to be an interpretable model.

topicLIME developed by \citeauthor{Kiefer2022} (\citeyear{Kiefer2022}) generates a local neighborhood of a document to be explained by removing coherent words. It is therefore capable of including distributional, contextual, as well as semantic information of the input domain in the resulting topic-based explanations. Doing so, it offers realistic and meaningful local perturbation distributions by avoiding extrapolation when generating the local neighborhood, leading to higher local fidelity of the local surrogate models. For a sample topic-based explanation, please refer to \citep{Kiefer2022}.

\subsection{Our Method: SemanticPush} \label{SemanticPush}
Our method called SemanticPush shall enable model-agnostic Interactive Machine Learning on a higher level of semantic detail. It therefore extends the idea of CAIPI (refer to the CAIPI algorithm in the Appendix) that offers humans model-agnostic, albeit contextless interactions in the form of word-based explanations and 'destructive' corrections.
From IML research it is known that humans want to demonstrate how learners \textit{should} behave. According to \citeauthor{Amershi2014} (\citeyear{Amershi2014}) and \citeauthor{Odom2018} (\citeyear{Odom2018}), people don`t want to simply teach 'by feedback', but want to teach 'by demonstration' or by providing examples of a concept. Therefore, interaction techniques should move away from limited, learner-centered ways of interactions, but rather proceed to more natural feedback, such as suggesting alternative or new features \citep{Stumpf2007}.
\begin{figure}[ht!]
	\centering
	\includegraphics[width=0.25\textwidth]{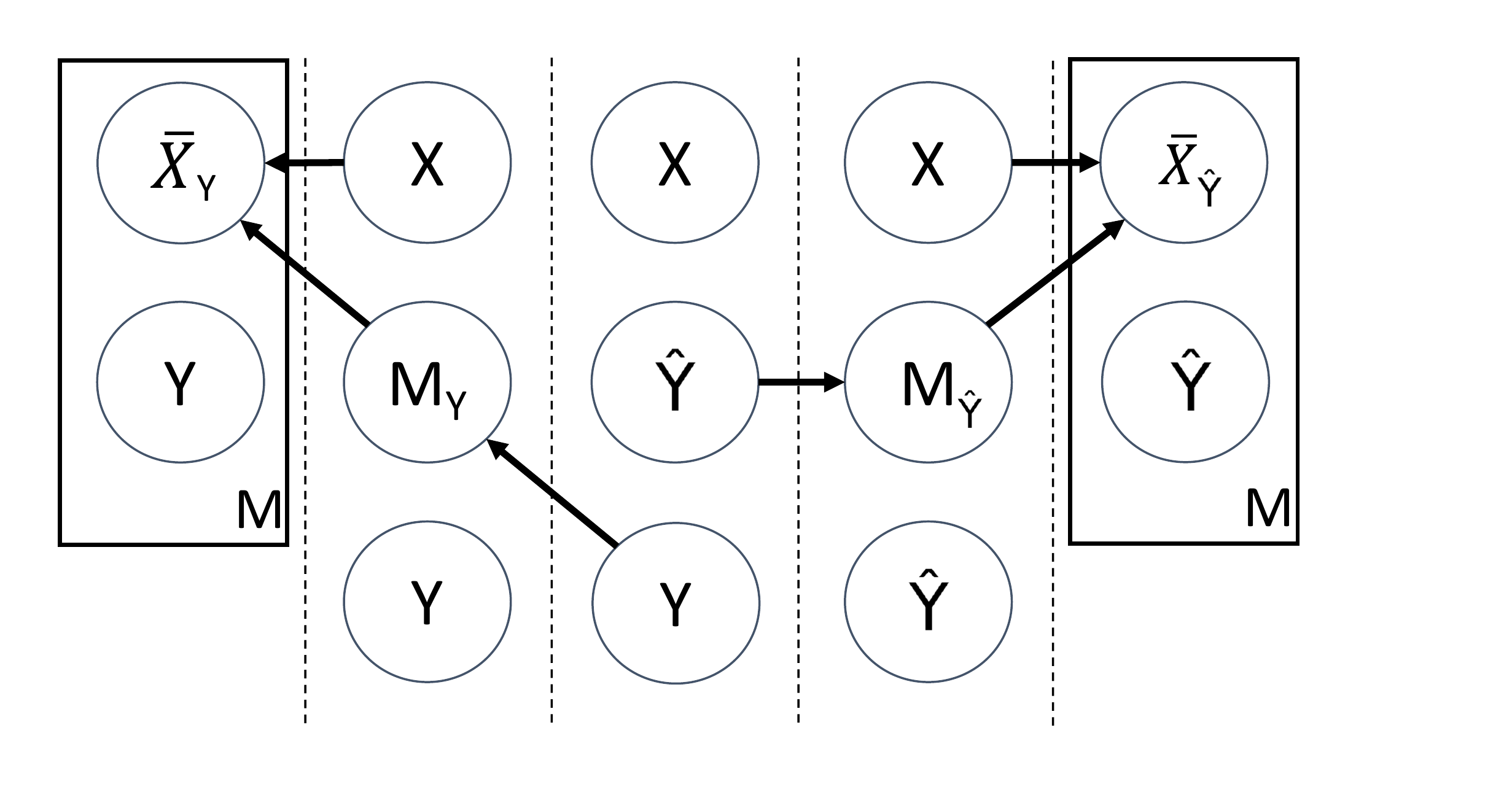}
	\caption{Graphical Model of SemanticPush.}
	\label{img:GraphicalModel}
\end{figure}

SemanticPush shall put that knowledge into practice and is depicted in figure \ref{img:GraphicalModel} as graphical model.
Let $X$ and $Y$ be the input and output space for a binary classification, where $X$ represents query instances, $Y$ and $\hat{Y}$ true and predicted labels respectively.
The overall goal is to find a matrix $M$ depending on the label $Y$ that adequately incorporates human feedback into the classifier's reasoning in a model-agnostic way by generating counterexamples $\bar{X}$ based on $X$.
Thus, we are seeking for a set of $L$ input manipulations $M=\{m_1, ..., m_L\}$ as well as a manipulation function $q:M \times X \to \bar{X}$.
$q(m,x)$ shall be a local function such that it only affects a part of the input $X$. This is the case because user input in IML shall be focused (it shall only affect a certain part/aspect of the model) as well as incremental (each user input shall only result in a small change of the model) \citep{Amershi2014}.

Algorithms \ref{alg:Semantic Push} and \ref{alg:Semantic Correction} describe SemanticPush in detail.

\begin{algorithm}[ht!]
\begin{algorithmic}
\Require a destructive correction set $C_{{dest}_x}$, a topicLIME explanation $\hat{z}_{xy}$ for query instance $x$ with true class $y$, expert knowledge - here simulated via Gold Standard $GS$ - and a balancing parameter $\lambda$
\State $C_{{dest}_x}= \{t \in \hat{z}_{xy} \vert t \notin GS_y \}$

\If{$\hat{y} = y \land C_{{dest}_x} \neq \emptyset$}
	\Comment Right for the partially wrong reasons
 	\State $\bar{x}_i \gets x \setminus  C_{{dest}_x}  \cup$ \textcolor{SeaGreen}{$\textbf{S}emantic\,\textbf{C}ompletion$}(\\\ $ x, GS_y, \hat{z}_{xy}, \lambda$)
	\Comment Add a concept the classifier forgot to learn
	\State $\bar{y} \gets y$
\ElsIf{$\hat{y} \neq y$}
	\Comment False prediction
	\State $\bar{x}_{iy} \gets$ \textcolor{red}{$\textbf{S}emantic\,\textbf{C}orrection$}$_y$($x, GS_y, \hat{z}_{xy}$)
	\Comment Provide feedback/hints for the true class
	\State $\bar{y}_{iy} \gets y$
	\State $\bar{x}_{i\hat{y}} \gets$ \textcolor{red}{$\textbf{S}emantic\,\textbf{C}orrection$}$_{\hat{y}}$($x, GS_{\hat{y}}, \hat{z}_{x\hat{y}}$)
	\Comment Provide feedback/hints for the predicted class
	\State $\bar{y}_{i\hat{y}} \gets \hat{y}$
\EndIf
\end{algorithmic}
\caption{SemanticPush}
\label{alg:Semantic Push}
\end{algorithm}

\begin{algorithm}[ht!]
\begin{algorithmic}
\Require a Topic Model $lda$
\State $\theta_x \gets$ $lda$.$\textbf{G}et\,\textbf{T}opic \, \textbf{M}ixture$(x)
\For{$t\in \theta$}
	\Comment $t$ represents a topic as explanation unit
	\State
	\begin{varwidth}[t]{\linewidth}
	\If {$t \in \hat{z}^+ \cap GS^+ \lor t \in \hat{z}^- \cap GS^- \lor$ \par
        \hskip\algorithmicindent$ t \in \hat{z}^+ \cap GS^-  \lor (t \notin \hat{z} \land t \notin GS)$}
        \Comment Topics either correctly used or incorrectly used (but hard to reverse polarity and still important) or correctly ignored
		\State $\hat{\theta}_{x_t} \gets$ KeepProbability($\theta_{x_t}$)
	\ElsIf{$t \in \hat{z}^- \cap GS^+ \lor$ \par
        \hskip\algorithmicindent $(t \notin \hat{z} \land t \in GS^+)$}
        \Comment Topics either incorrectly learned (but easy to reverse polarity) or forgotten to learn
		\State $\hat{\theta}_{x_t} \gets$ \textcolor{Turquoise}{$\textbf{I}ncrease\,\textbf{P}robability$}($\theta_{x_t}, GS, \lambda$)
	\ElsIf{$(t \in \hat{z} \land t \notin GS)$}
		\Comment Irrelevant topics were used
		\State $\hat{\theta}_{x_t} \gets$ \textcolor{Goldenrod}{$\textbf{D}ecrease\,\textbf{P}robability$}($\theta_{x_t}$)
	\EndIf
\end{varwidth}
\EndFor
\State
\Return $lda$.$\textbf{S}ample\,\textbf{I}nstance$($\psi(\hat{\theta}_x)$)
\Comment sampling from the multinomial distribution harnessing the generative process of LDA 
\end{algorithmic}
\caption{\textcolor{red}{$\textbf{S}emantic\,\textbf{C}orrection$}}
\label{alg:Semantic Correction}
\end{algorithm}

\textcolor{SeaGreen}{$\textbf{S}emantic\,\textbf{C}ompletion$}($x, GS_y, \hat{z}_{xy}, \lambda$) from algorithm \ref{alg:Semantic Push} is defined as  $\sim [\lambda * \psi(C_{{add}_x}) + (1-\lambda) * \psi(x_{add})]$,
where
$C_{{add}_x} = \{(t, t_w) \in GS_y^+ \vert t \notin \hat{z}_{xy}^+\}$
 and
$x_{add} = \{(t, t_w) \in x \vert t \in C_{{add}_x} \}$.

$\psi$ constitutes a normalization operator that re-normalizes the weights $t_w$ of the according topics $t$ (be it from Gold Standard or topicLIME explanation), revealing a multinomial distribution over topics $t$. SemanticPush then incorporates the concepts the classifier forgot to learn by adding text parts via sampling ($\sim$) from the multinomial distribution and harnessing the generative process of LDA (see subsection \ref{LDA}).

\textcolor{Turquoise}{$\textbf{I}ncrease\,\textbf{P}robability$}() from algorithm \ref{alg:Semantic Correction} performs a topic`s probability change $\delta_t$ in the following way:\\
$\delta_t = \theta_{x_t} + \lambda * GS_{y_t} + (1 - \lambda) * \theta_{x_t}$.

\textcolor{Goldenrod}{$\textbf{D}ecrease\,\textbf{P}robability$}() from algorithm \ref{alg:Semantic Correction} in our scenario sets a topic's probability to zero as the topic is assumed to be irrelevant for the class decision.

\begin{figure}[ht!]
	\subfigure [] {\includegraphics[width=0.41\textwidth]{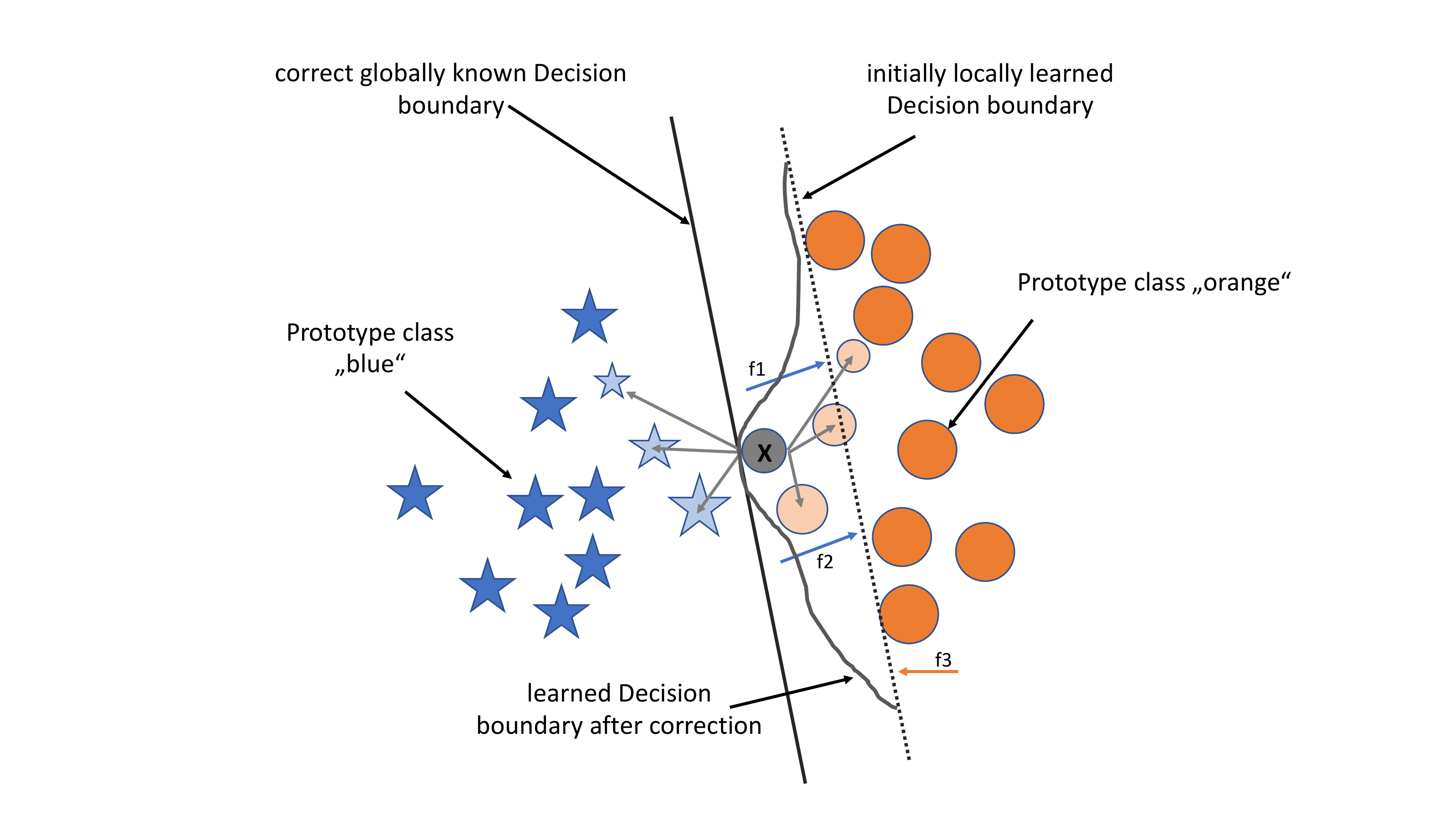}}
	\subfigure [] {\includegraphics[width=0.41\textwidth]{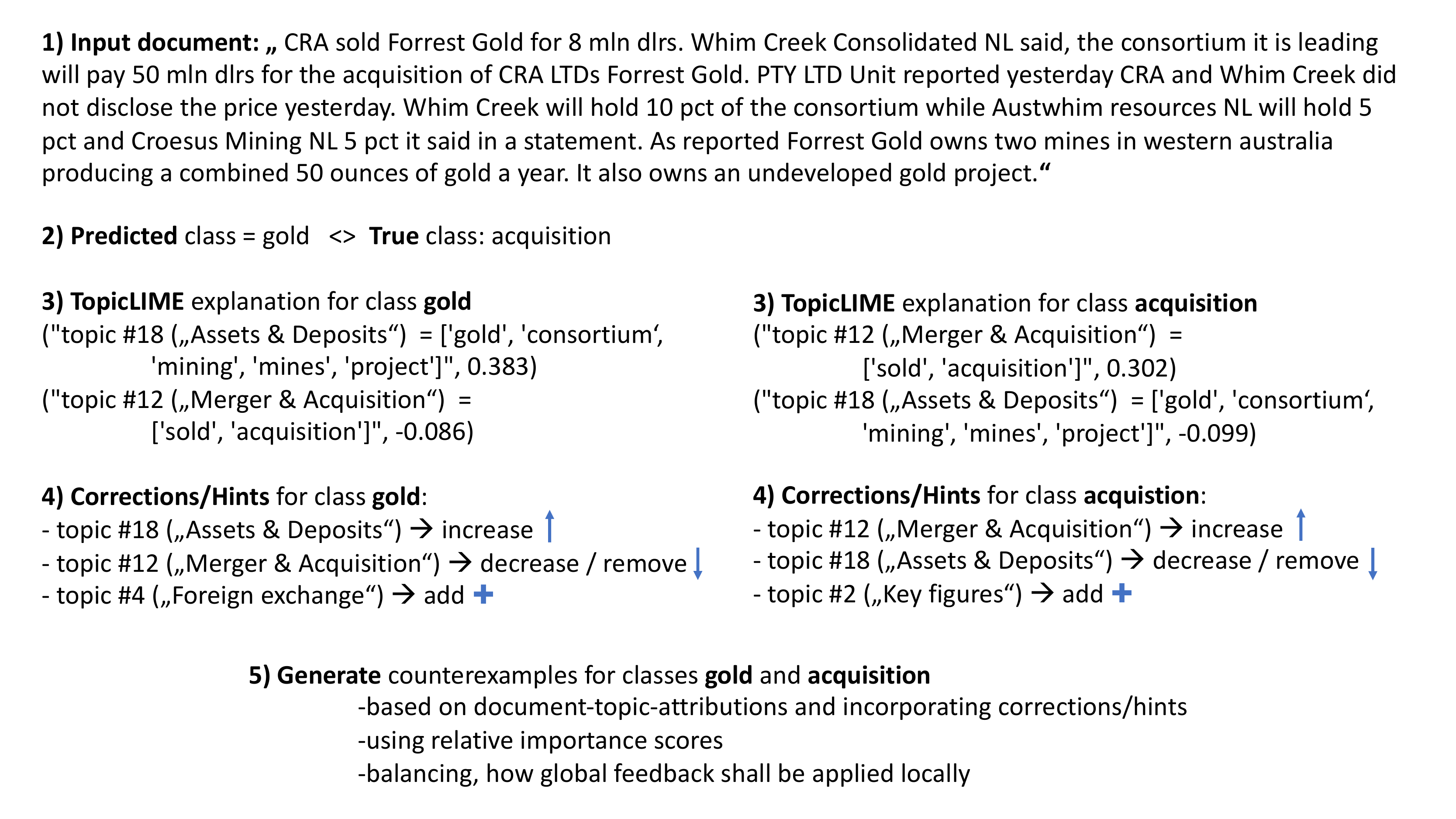}}
	\caption{(a) Conceptualization of SemanticPush: The grey query instance in the middle is predicted as class 'blue', but should be 'orange' instead according to ground truth. Local explanation features f1 and f2 are features used by the classifier locally to assign the query instance to class 'blue'. According to expert knowledge, those features push the learned local decision boundary too far towards the class 'orange'. Feature f3 also constitutes expert knowledge as it is, among others, significantly used globally by the classifier to assign instances to class 'orange'. SemanticPush incorporates this information by generating new instances (shown in light color) for both classes and eventually weighs them by their distance to the query instance. The degree of locality of applying the expert knowledge to the query instance is controlled by the hyperparameter $\lambda$. Sampling new instances only based on global expert knowledge might result in prototypical instances (located in dense regions) which might not lead to great benefit for the classifier. (b) An examplary application of SemanticPush to document ID 9 of the Reuters R 52 Dataset.}
	\label{img:SemanticPush}
\end{figure}

SemanticPush is based on a so-called \textit{conceptual Gold Standard} $GS$ that works as a proxy for an oracle`s expert knowledge. Specifically, $GS_y$ contains concepts in the form of LDA-retrieved topics that should be informative for a specific class $y$.
We reveal that kind of Gold Standard using intrinsic feature selection, especially by extracting the weights of a Logistic Regression Model trained on all available topic-represented data from the datasets described in subsection \ref{Datasets}. Details on how we implemented the $GS$ can be found in subsection \ref{Models}. The superscripts $+$ and $-$ (of Gold Standard $GS$ or of the explanations $z$ respectively) indicate positive and negative attributions for a specific class.
In addition to the algorithmic descriptions, figures \ref{img:SemanticPush}(a) and \ref{img:SemanticPush}(b) illustrate SemanticPush conceptually and with an exemplary application.

SemanticPush differs from CAIPI (a) in using contextual topicLIME explanations instead of LIME explanations, (b) in internally using a conceptually meaningful Gold Standard that allows corrections on higher semantic detail, (c) in additionally enabling constructive feedback, and (d) in also being able to locally correct the reasoning for false predictions.
Transferred to a real-world interaction setting, human annotators are not only capable of indicating and correcting (a) components that a learner wrongly identified as relevant (as CAIPI does), but also (b) components that have been forgotten to learn, and (c) components which are relevant but have been incorrectly used.

\section{Experimental Setup}
\subsection{Baseline: Active Learning and CAIPI}
We compare our SemanticPush approach against three baseline approaches: First, we use a standard ActiveLearner that internally harnesses Maximum Classification Uncertainty with regard to a pool dataset as sampling strategy. Classification uncertainty is defined as $U(x) = 1- P_\theta(\hat{y} \vert x) $, where $x$ is the instance to be predicted and $\hat{y}$ is the most likely prediction. 
Second, we apply the original CAIPI method as described in \citep{Teso2019} that provides explanation corrections for the 'right for the wrong reasons' \mbox{($\hat{y}=y$)} case. We call the according setup 'CAIPI destructive' (CAIPI$_{d}$) as it is only capable of removing those components that have been identified by a local LIME explanation $\epsilon(x)$  as relevant, but an oracle believes those components to be irrelevant.
Third, we extend CAIPI such that it is additionally able to deal with false predictions ($\hat{y} \neq y$). We call that setting 'CAIPI destructive + constructive' (CAIPI$_{d/c}$) as we additionally generate new documents comprising words that could have been used to predict the according true class. We therefore sample words from a set $GS_{local}^+(x)$ (where
$GS_{local}^+(x) = GS_{global}^{(k^+)}(y) \cap x$) that contains the top $k$ positive words from a global Gold Standard of the true class (see subsection \ref{SemanticPush}) that are also part of the document.

\subsection{Datasets} \label{Datasets}
We evaluate SemanticPush on two multiclass classification tasks harnessing the following datasets: \textit{AG News Classification} Dataset \citep{Xiang2015} and \textit{Reuters R52} Dataset \citep{Lewis1993}. The \textit{AG News} Dataset (127,600 documents) is constructed by selecting the four largest classes from the original AG Dataset, which is a collection of more than one million news articles. The average document length is 25 words, classes to be distinguished are 'Business News', 'Science-Technology News', 'Sports News', and 'World News'.
The \textit{Reuters R52} Dataset (9,100 documents) originally comprises 52 classes. Due to strong imbalance between the classes, we selected the ten most represented classes ('Earn', 'Acquisition', 'Coffee', 'Sugar', 'Trade', 'Ship', 'Crude', 'Interest', and 'Money-Foreign-Exchange'), leading to a corpus comprising 7,857 documents. The average document length is 60 words. From now on, we refer to this dataset as the \textit{Reuters R10} Dataset.

For both datasets, we perform standard NLP preprocessing steps like Tokenization, Lemmatization, Stemming, Lower-Casing, and Removing of stopwords.
\subsection{Models} \label{Models}
Our architecture comprises a semantic component that provides contextual information about the input domain. Here, we showcase how we instantiated the \textbf{Latent Dirichlet Models} for the two datasets.
Throughout this research, we used scikit-learn (version 0.20.2) and gensim (version 3.8.3).
For the \textit{AG News} Dataset, several LDA models have been trained on the preprocessed corpus with different values for the hyperparameter \textit{number of topics k}. A final selection has been made by determining the optimal number  $K^*$ of topics $t = 1,...,K$ by solving $\scriptsize \underset{K}{\arg\max} \ \frac{1}{K} \sum \limits_{t=1}^K C\textsubscript{v}(t)$, where $C\textsubscript{v}$ is the C\textsubscript{v} coherence as introduced in subsection \ref{LDA}. We set $K$ to 30 and determined $K^* = 13$, meaning an optimal number of 13 topics.
Those topics, together with its most representative words, are described in figure \ref{img:LDA_topics} in the Appendix.

Analogously, we proceeded with the \textit{Reuters R10} Dataset, but in contrast to the \textit{AG News} Dataset we could not solely rely on C\textsubscript{v} coherence to find a suitable number of topics. As the LDA model in our framework not only serves as semantic component, but is also used to build a topic-based Gold Standard model (see next paragraph), we rather had to trade off C\textsubscript{v} coherence against learning performance. In order to achieve sufficient predictive performance for \textit{Reuters R10} while preserving high coherence, the optimal number of topics $K^*$ was set to 100.

\textbf{Gold Standard} 
As described in subsection \ref{SemanticPush}, a Logistic Regression model is harnessed as an approximation for the oracle's expert knowledge required in any Active Learning setting. To obtain that kind of Gold Standard $GS$ for CAIPI, we trained the regression model on the bag-of-words-represented documents and got the following results:

For the \textit{AG News} Dataset, a macro-averaged F1 score of 0.85 was achieved, for \textit{Reuters R10}, the regression model reached a score of 0.8.

In order to include contextual and higher-level semantic information (simulating conceptual knowledge of a human expert) in the $GS$ used for SemanticPush, we represented the documents as multinomial distributions over topics (features of the regression model) using the LDA model described above. The according model achieved a macro-averaged F1 score of 0.74 for \textit{AG News} and of 0.71 for \textit{Reuters R10}. Due to the reduced number of features when representing documents via topics, the topic-based $GS$ obviously performs slightly worse than the word-based $GS$ due to reduced degrees of freedom of the regression model.

%
%
%
%
%

During our experiments, we primarily used an XGBoost model as \textbf{Base Learner} as it constitutes a high-performing ensemble and tree-based classification algorithm that can intrinsically learn feature interactions. In addition, we also experimented with a Support Vector Machine with linear kernel. For instantiating the Active Learner, we chose the modAL python framework \citep{Danka2018}. As query strategy, we used Maximum Classification Uncertainty. For both datasets, a stratified split into train-, pool-, and testsets was performed (train 1\%, pool 79\%, and test 20\% of the data). We therefore account for a standard Active Learning setting where only a small number of labeled data, but a huge number of unlabeled data is available. All experiments were performed over 200 iterations each.

\subsection{Evaluation Metrics}
For evaluating the quality of our framework and for answering the research questions two and three (see section \ref{Introduction}), we performed two kinds of experiments. 
First, we measured the \textbf{Predictive Performance} of the different IML strategies with regard to a downstream classification task on the testset during 200 iterations. As performance metric for evaluation of research question two, we chose the macro-averaged F1 score (after each AL iteration) on the one hand and the Average Classification Margin between predicted and true class (after every tenth AL iteration) on the other hand. The Average Classification Margin between predicted and true class is defined as $\scriptsize M(x) = \frac{1}{N} \sum \limits_{i=1}^N P(\hat{y} \vert x_i) - P(y \vert x_i)$, where $\hat{y}$ is the predicted class and $y$ is the true class, $\hat{x_i}$ is a certain instance of the testset to be predicted, and $N$ is the total number of instances in the testset. Accordingly, this measure analyzes the classifier's confidence towards false predictions for all test instances and then averages over those.

To answer research question three, \textbf{Local Explanation Quality} was analyzed two-fold: (a) with regard to local fidelity and approximation accuracy (the quality of the local explanation generators itself before any interactions) and (b) with regard to the 'Explanation Ground Truth' of the downstream classification tasks (the quality of local explanations for all test instances compared to the bag-of-words-represented Gold Standard described in subsection \ref{Models}).

(a): Local fidelity is said to be achieved if an explanation model $g$ $\epsilon$ $G$ is found such that $f(z)$ $\approx$ $g(z')$ for $z, z'$ $\epsilon$ $Z$, where $Z$ constitutes the vicinity of $x$ and $f$ is the model to be explained.
We use Mean Local Approximation Error (MLAE, equation \ref{eq:2}) and Mean$R^{2}$ (equation \ref{eq:3}) as a proxy to measure local fidelity of the whole explanation models to be compared.

\begin{equation} \label{eq:2}
\scriptsize
MLAE = \frac{\sum \limits_{i=1}^N {|f(x_i) - g_i(x_i)|}}{N}.
\end{equation}

\begin{equation} \label{eq:3}
\scriptsize
 MeanR^{2} = \frac{\sum \limits_{i=1}^N {R^{2}(g_i)}}{N},   R^{2} = 1 - \frac{\frac{1}{n} \sum \limits_{i=1}^n{(f(z_i) - g(z'_i))^2}}{\frac{1}{n} \sum \limits_{i=1}^n{(f(z_i) - f_{mean})^2}}.
\end{equation}

In both cases,  $N$ is the number of instances in the according test dataset.


Furthermore, a modified variant of the \textit{Area Over The Perturbation Curve} (AOPC) was analyzed. It measures local fidelity of individual explanations, we call it Combined Removal Impact ($CRI$) and define it as:

\begin{equation} \label{eq:4}
\scriptsize
CRI = \frac{1}{N}\sum \limits_{i=1}^N {p(\hat{y}|x_i) - p(\hat{y}|\tilde{x}_i^{(k)})},
\end{equation}

where the top \textit{k\%} explanation features are removed from $x_i$ yielding $\tilde{x}_i^{(k)}$, $\hat{y}$ denotes the predicted label for $x_i$ and $N$ is the number of instances in the according test dataset.
For both evaluation metrics, please refer to \citep{Kiefer2022} in order to find details on how those metrics have been applied to compare word-based and topic-based contextual explanations.

(b): In order to analyze the development of Local Explanation Quality after applying the different IML strategies, we calculated a measure called 'Explanatory Accuracy'.
First, we took $k=10\%$ of the most relevant words from global Gold Standard $GS_{global}^{(k)}(y)$ and intersected those with a document's words ($GS_{global}^{(k)}(y) \cap x$)  resulting in a local Gold Standard ($GS_{local}(x)$). Subsequently, for each test document $x$ a local explanation $\epsilon(x)$ was generated using LIME. The Average Explanatory Accuracy  was then defined as:

\begin{equation} \label{eq:5}
\scriptsize Explanatory Accuracy_{AVG} = \frac{1}{N} \sum \limits_{i=1}^N \frac{\vert GS_{local}(x_i) \cap \epsilon(x_i) \vert}{\vert GS_{local}(x_i) \vert}
\end{equation},
 with $N$ being the number of documents in the test dataset. We restricted the complexity of the local surrogate models (number of explanatory words) to $\Omega(g) = \vert GS_{local}(x)  \vert$, such that the LIME explanations were theoretically capable of finding all relevant explanations according to local $GS$. We measured the Average Explanatory Accuracy of the test instances after every 20th iteration.

\section{Experiment 1: Predictive Performance}
\begin{figure}[ht!]
	\subfigure [] {\includegraphics[width=0.35\textwidth]{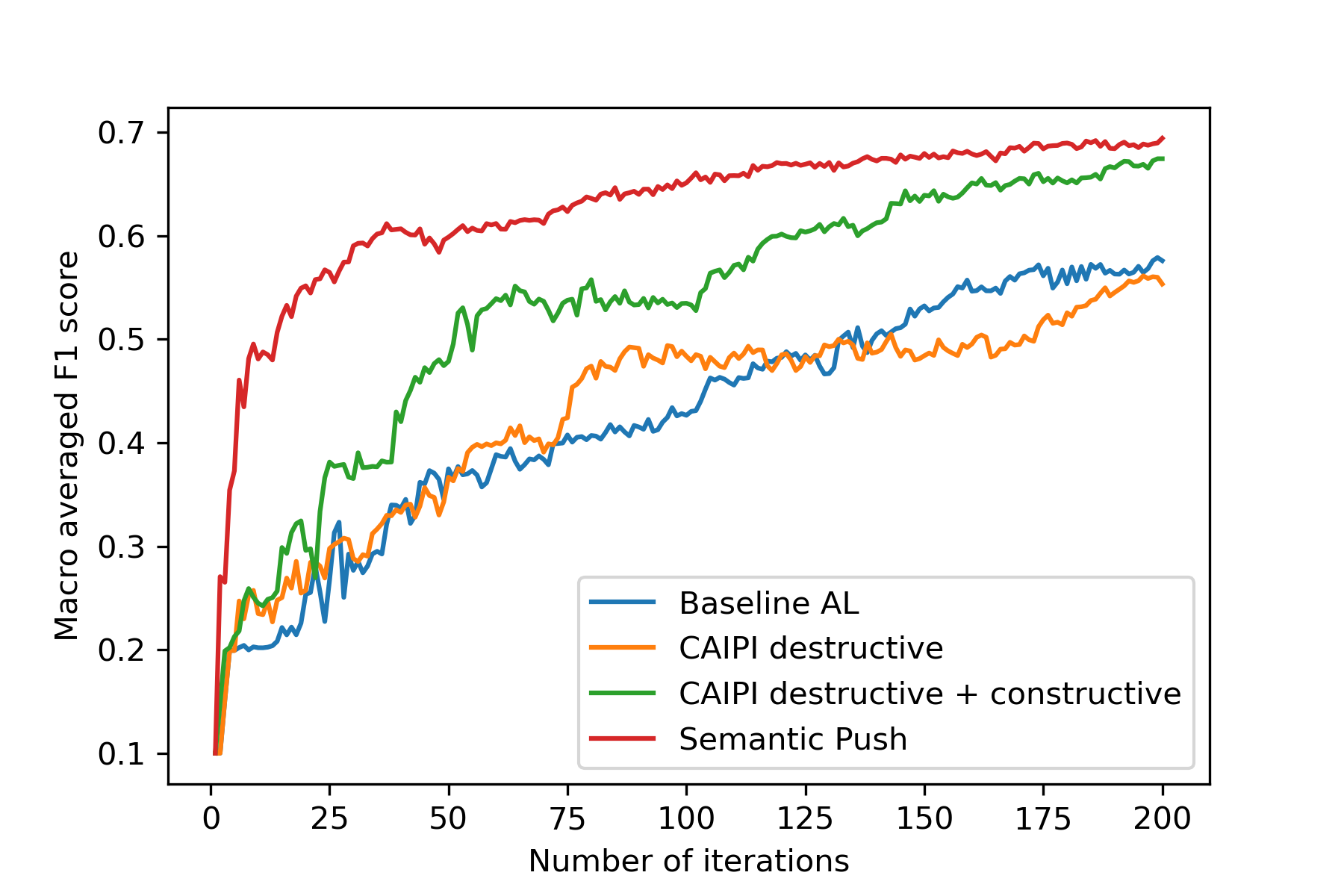}}
	\subfigure [] {\includegraphics[width=0.35\textwidth]{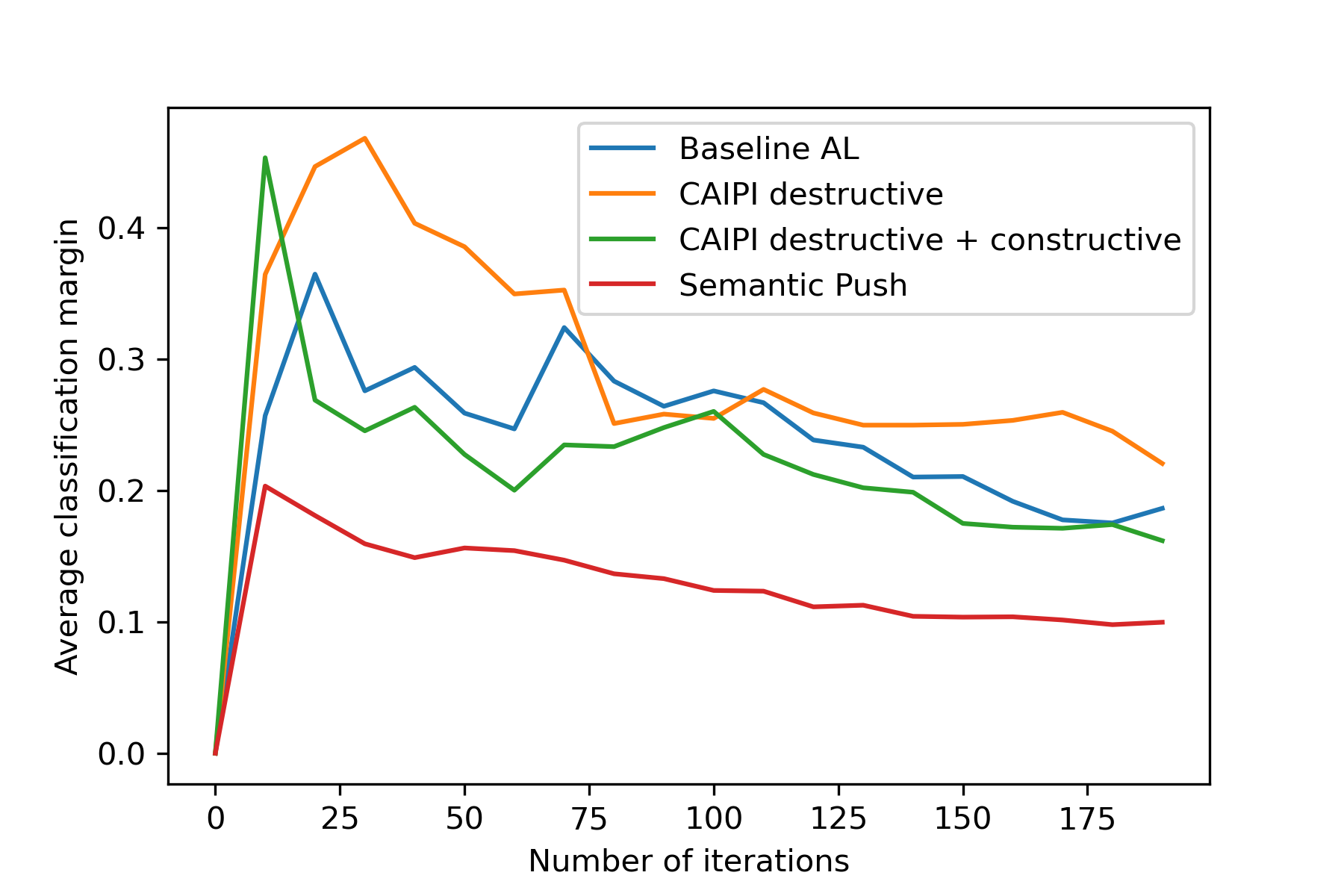}}
	\caption{(a) Learning performance of different IML strategies for \textit{AG News} Dataset. (b) Average Classification Margin of different IML strategies for \textit{AG News} Dataset.}
	\label{img:PerformanceAndMargin_AG}
\end{figure}

\begin{figure}[ht!]
	\subfigure [] {\includegraphics[width=0.35\textwidth]{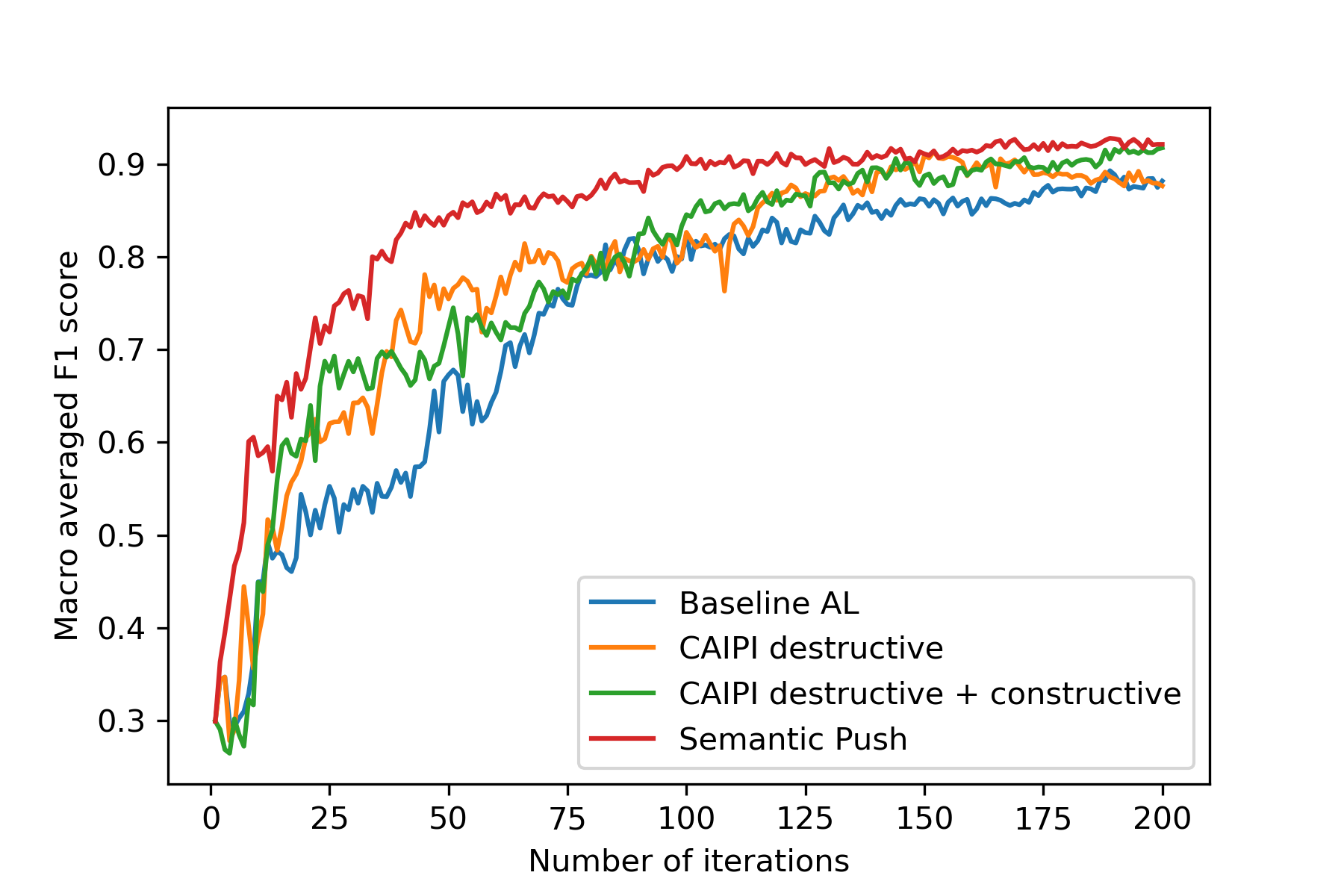}}
	\subfigure [] {\includegraphics[width=0.35\textwidth]{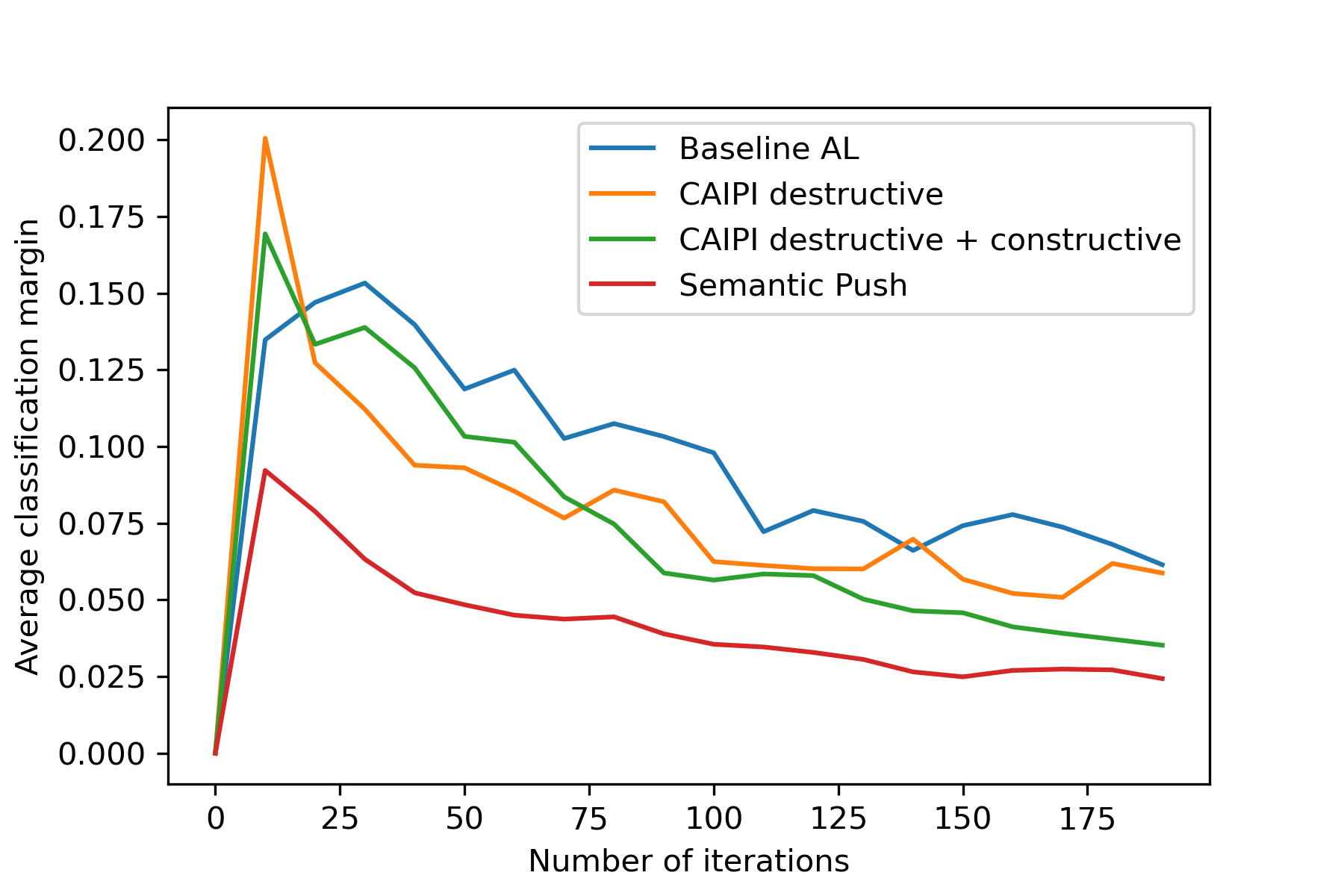}}
	\caption{(a) Learning performance of different IML strategies for Reuters R10 Dataset. (b) Average Classification Margin of different IML strategies for Reuters R10 Dataset.}
	\label{img:PerformanceAndMargin_Reuters}
\end{figure}

We conducted the first experiment by measuring the \textbf{Predictive Performance} of the different IML strategies. 
Figures \ref{img:PerformanceAndMargin_AG}(a) and \ref{img:PerformanceAndMargin_Reuters}(a) show the convergence of the macro averaged F1 score on the two testsets over 200 iterations for our approach SemanticPush as well its baselines. For both datasets, SemanticPush clearly outperforms the standard ActiveLearner and the two versions of CAIPI when using XGBoost as base classifier (despite a Gold Standard that is around ten percent worse than the one used for CAIPI). 
It can also be seen that SemanticPush incorporates the oracle's expert knowledge efficiently at much earlier stage (around 90 percent of final F1 score reached already after only 50 iterations). In the middle range of the iterations, SemanticPush has already applied much of the correct knowledge and therefore its performance starts to increase more slowly. For classifiers like the Support Vector Machine (see figure \ref{img:PerformanceAndMargin_Reuters_SVC}(a) in the Appendix), which earlier reached high classification accuracy (in the realm of the conceptual Gold Standard's performance), SemanticPush's performance starts to stagnate during later iterations as it partially has applied 'incorrect corrections'.
\textit{CAIPI destructive} is not able to consistently beat the ActiveLearner's baseline, while our constructive extension performs better. 
Figures \ref{img:PerformanceAndMargin_AG}(b) and \ref{img:PerformanceAndMargin_Reuters}(b) confirm those observations from the point of the Average Classification Margin between predicted and true class, where SemanticPush on average performs false predictions less frequent or with less confidence than its baselines.
Across all experiments, we kept the hyperparameters constant. At each iteration, we allowed the different methods to generate $M=10$ counterexamples incorporating the corrective knowledge. Furthermore, we set the length (number of words) of each counterexample to the average document length of the respective corpora (25 for the \textit{AG News} Dataset and 60 for \textit{Reuters R10}).
We allowed LIME to generate explanations containing 7 words (for \textit{AG News}) and 15 words (for \textit{Reuters R10).}
The topicLIME explanations included 3 and 5 topics respectively. This limitation was made due to the fact that in real world humans can only perceive, process and remember a limited number of information. According to Miller's law \citep{Miller1956}, this capacity is somewhere between seven plus or minus two. Additionally, we set $\lambda$ to 0.95 as we simulated some sort of global expert knowledge.

\subsection{Experiment 2: Local Explanation Quality}
We performed experiments by analyzing the \textbf{Local Explanation Quality} of the different IML strategies in both directions of interaction with the oracle. Table \ref{tbl:localfid} compares the quality of the local surrogate models and the resulting explanations generated by LIME and topicLIME. The related measures are Approximation Error,  $MeanR^{2}$ as well as Combined Removal Impact (CRI) of the two different test datasets. It is noticeable that both the surrogate explanation models and the local explanations itself are more faithful towards the model to be explained when using contextual explanations generated from realistic local perturbation distributions. Therefore, resulting explanations are regarded as more reliable.
Table \ref{tbl:explqual} takes up the topic of Local Explanation Quality from the other direction (after the interactions with the oracle).
It is striking that only SemanticPush is capable of clearly incorporating the expert knowledge in a way that it is adequately adapted by the base classifier. The two versions of CAIPI don't reveal better results than the standard ActiveLearner.

\begin{table}[ht!]
\tabcolsep1.5mm
\caption{Comparison of LIME and topicLIME w.r.t. local fidelity} 
\centering 
\label{tbl:localfid}
\begin{tabular}{ccccccc} %
\toprule

 & \multirow{2}{*}{
\parbox[c]{.2\linewidth}{\centering}}
 & 
 \multicolumn{3}{c}{\textbf{AG News}} &&
\\

 & & {\centering \textbf{Lime}} & {\textbf{topicLime}} && {\textbf{Difference}}  \\
\midrule

& Approx. Error & 0.0394 & 0.0342 && -13\%\\
& $R^{2}$ & 0.863 & 0.884 && +2.5\% \\
& CRI & 0.229 & 0.277 && +21\% \\
\midrule
 && 
 \multicolumn{3}{c}{\textbf{Reuters R52}} &&\\
 \midrule 
& Approx. Error & 0.0195 & 0.0076 && -61\%\\
& $R^{2}$ & 0.864 & 0.951 && +10\% \\
& CRI & 0.271 & 0.302 && +11\% \\

\bottomrule
\end{tabular}
\end{table}

%
%
%
%
%
%

To sum it up, our proposed approach not only improves Learning Performance already at early stages of interactions, but also pushes the reasoning of the learner towards the desired behavior.

\begin{table}[ht!]
\tabcolsep1.5mm
\caption{Local explanation quality w.r.t. 'Ground Truth' of downstream classification tasks} 
\centering 
\label{tbl:explqual}
\begin{tabular}{cccccc} %
\toprule

   & \multirow{2}{*}{
\parbox[c]{.2\linewidth}{\centering}}
 & 
\multicolumn{3}{c}{\textbf{Explanatory Accuracy$_{AVG}$}} \\ 
\cmidrule{3-6} 

 & &  {\textbf{AL}} & {\centering \textbf{CAIPI$_{d}$}} & {\textbf{CAIPI$_{d/c}$}} & {\textbf{Sem.Push}}  \\
\midrule

& AG News & 0.690 & 0.683 & 0.685 & 0.711 \\
& Reuters R10 & 0.741 & 0.739 & 0.742 & 0.768 \\

\bottomrule
\end{tabular}
\end{table}

\section{Conclusion}
We introduced a novel IML architecture called Semantic Interactive Learning that helps to bring humans in the loop and allows for richer interactions. We instantiated it with SemanticPush, the first IML strategy enabling semantic and constructive corrections of a learner, also for false predictions. Our approach offers locally faithful and contextual explanations and, building on that, qualifies humans to provide conceptual corrections that can be considered graded and continuous. The corrections are in turn integrated into the learner's reasoning via non-extrapolating and contextual additional training instances. 
As a consequence of combining richer explanations with more extensive semantic corrections, our proposed interaction paradigm clearly outperforms its baselines\footnote{Please note that our constructive extension for CAIPI outperforms original CAIPI as well in most experiments w.r.t. Learning Performance.} with regard to learning performance as well as local explanation quality of downstream classification tasks in the majority of our experiments.
As the simulation of expert knowledge via a global Gold Standard is a crucial aspect of our architecture, we plan to improve the simulation's accuracy as well as evaluate its quality using Inter-rater reliability. Additionally, we intend to further include a language model like BERT into our architecture such that generated counterexamples are not only semantically meaningful, but also linguistically, especially syntactically correct.

\bibliography{bibliography_jabref}

\clearpage
\section*{Technical Appendix}
\begin{figure}[h!]
	\includegraphics[width=0.49\textwidth]{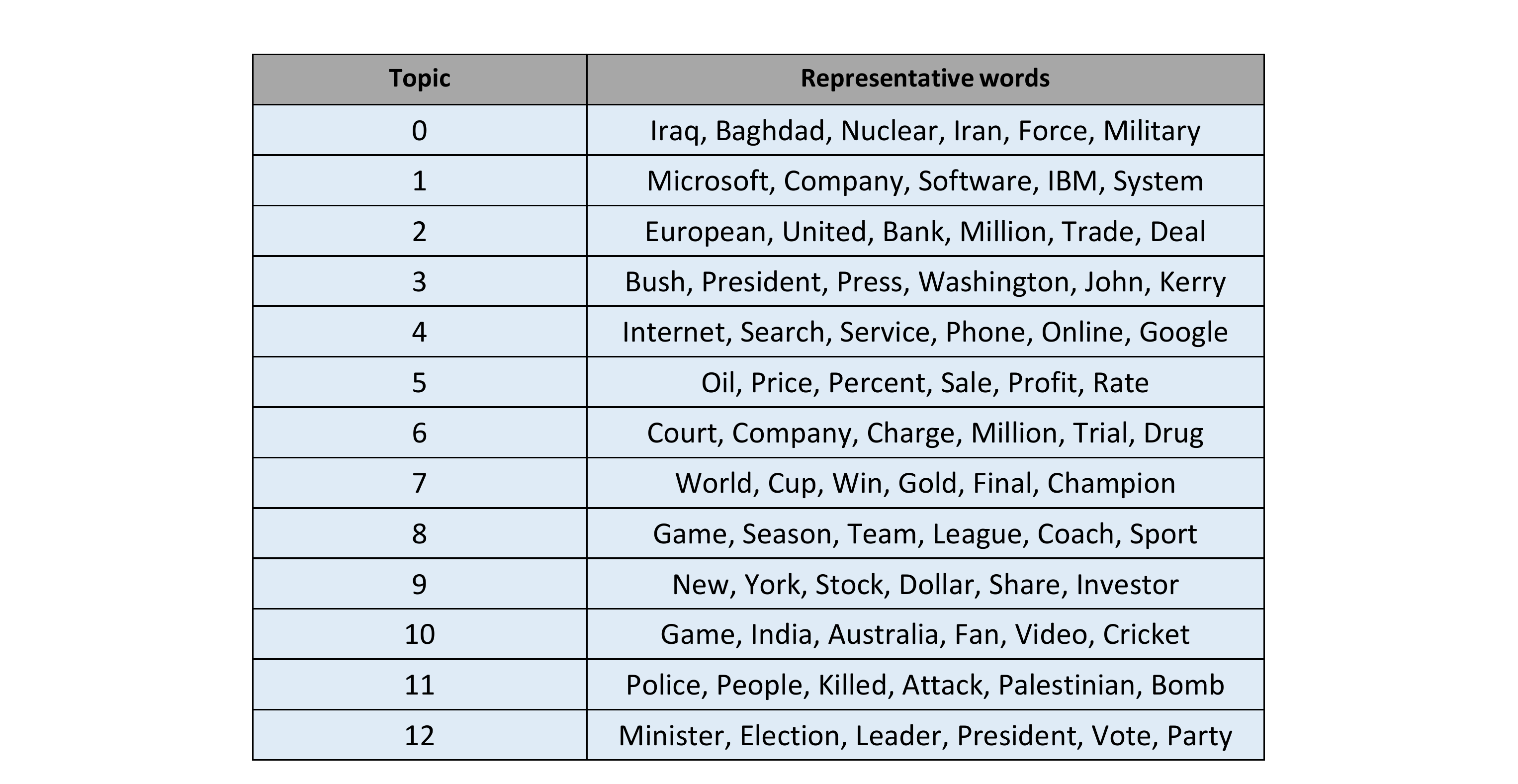}
	\caption{Learned LDA topics and its most representative words for  the \textit{AG News} Dataset}
	\label{img:LDA_topics}
\end{figure}

\begin{algorithm}[]
\begin{algorithmic}

\Require a set of labelled examples $L$, a set of unlabelled instances $U$ and an iteration budget $T$.
\State $f \gets FIT(L)$
\Repeat
	\State $x \gets$ \textcolor{Purple}{$\textbf{S}elect\,\textbf{Q}uery$}$(f,U)$
	\State $\hat{y} \gets f(x)$
	\State $\hat{z} \gets$  \textcolor{blue}{$\textbf{E}xplain$}$(f,x,\hat{y})$
	\State Present $x, \hat{y}$ and $\hat{z}$ to the user
	\State Obtain $y$ and explanation correction $C$
	\State $\{(\bar{x}_i, \bar{y}_i)\}_{i=1}^c \gets$ \textcolor{Orange}{$\textbf{T}o\,\textbf{C}ounterexamples$}$(C)$
	\State $L \gets L \cup \{(x,y)\} \cup \{\bar{x}_i, \bar{y}_i)\}_{i=1}^c$
	\State $U \gets U \setminus (\{x\} \cup \{\bar{x}_i\}_{i=1}^c)$
	\State $f \gets FIT(L)$
\Until budget $T$ is exhausted or $f$ is good enough
\State
\Return $f$
\end{algorithmic}
\caption{CAIPI \citep{Teso2019}}
\label{alg:CAIPI}
\end{algorithm}

\begin{table}[h!]
\tabcolsep1.5mm
\caption{Local explanation quality w.r.t. 'Ground Truth' of downstream classification task (Support Vector Machine)} 
\centering 
\label{tbl:explqual_SVC}
\begin{tabular}{cccccc} %
\toprule

   & \multirow{2}{*}{
\parbox[c]{.2\linewidth}{\centering}}
 & 
\multicolumn{3}{c}{\textbf{Explanatory Accuracy$_{AVG}$}} \\ 
\cmidrule{3-6} 

 & &  {\textbf{AL}} & {\centering \textbf{CAIPI$_{d}$}} & {\textbf{CAIPI$_{d/c}$}} & {\textbf{Sem.Push}}  \\
\midrule

& Reuters R10 & 0.786 & 0.785 & 0.788 & 0.796 \\

\bottomrule
\end{tabular}
\end{table}

\begin{figure}[]
	\subfigure [] {\includegraphics[width=0.5\textwidth]{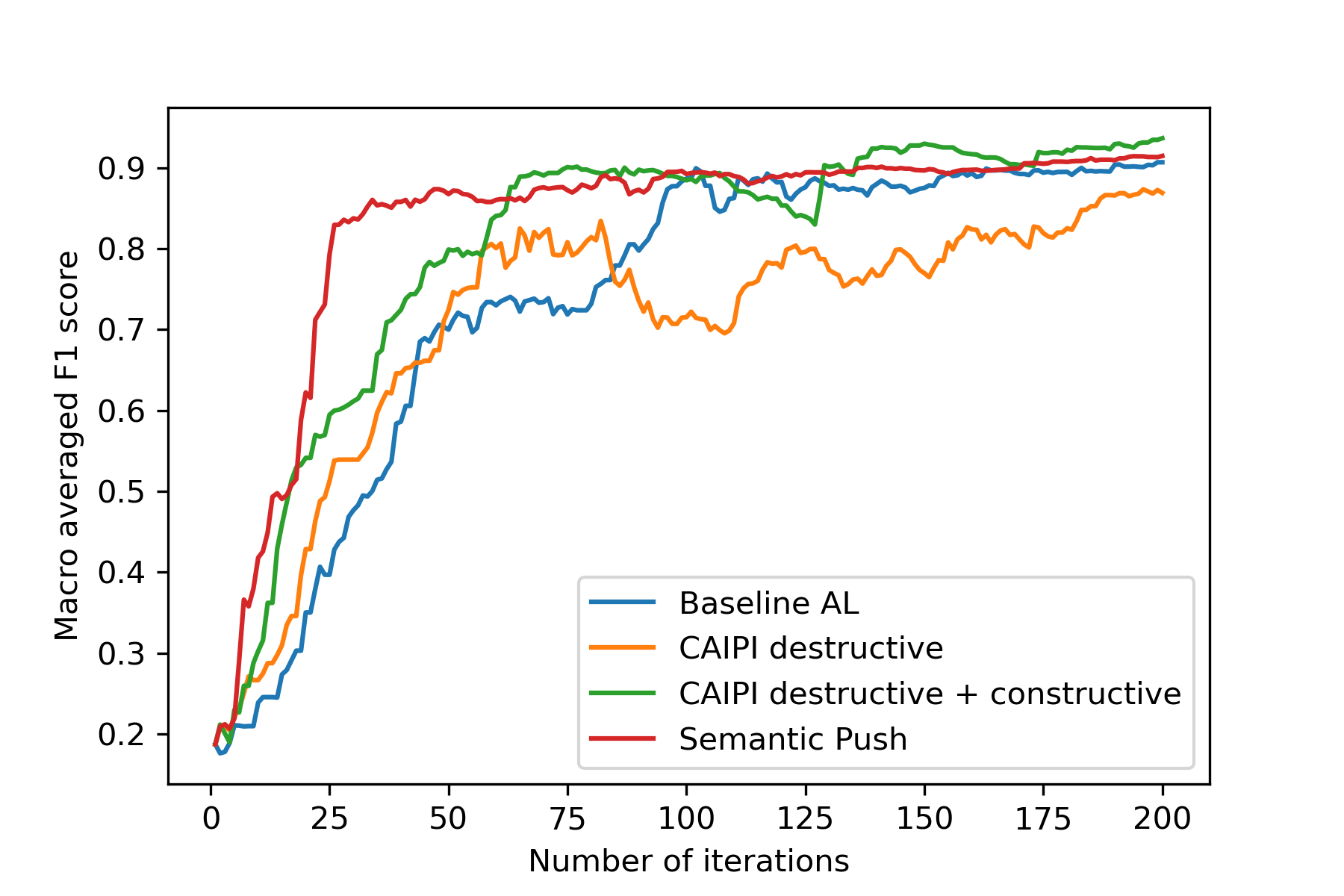}}
	\subfigure [] {\includegraphics[width=0.5\textwidth]{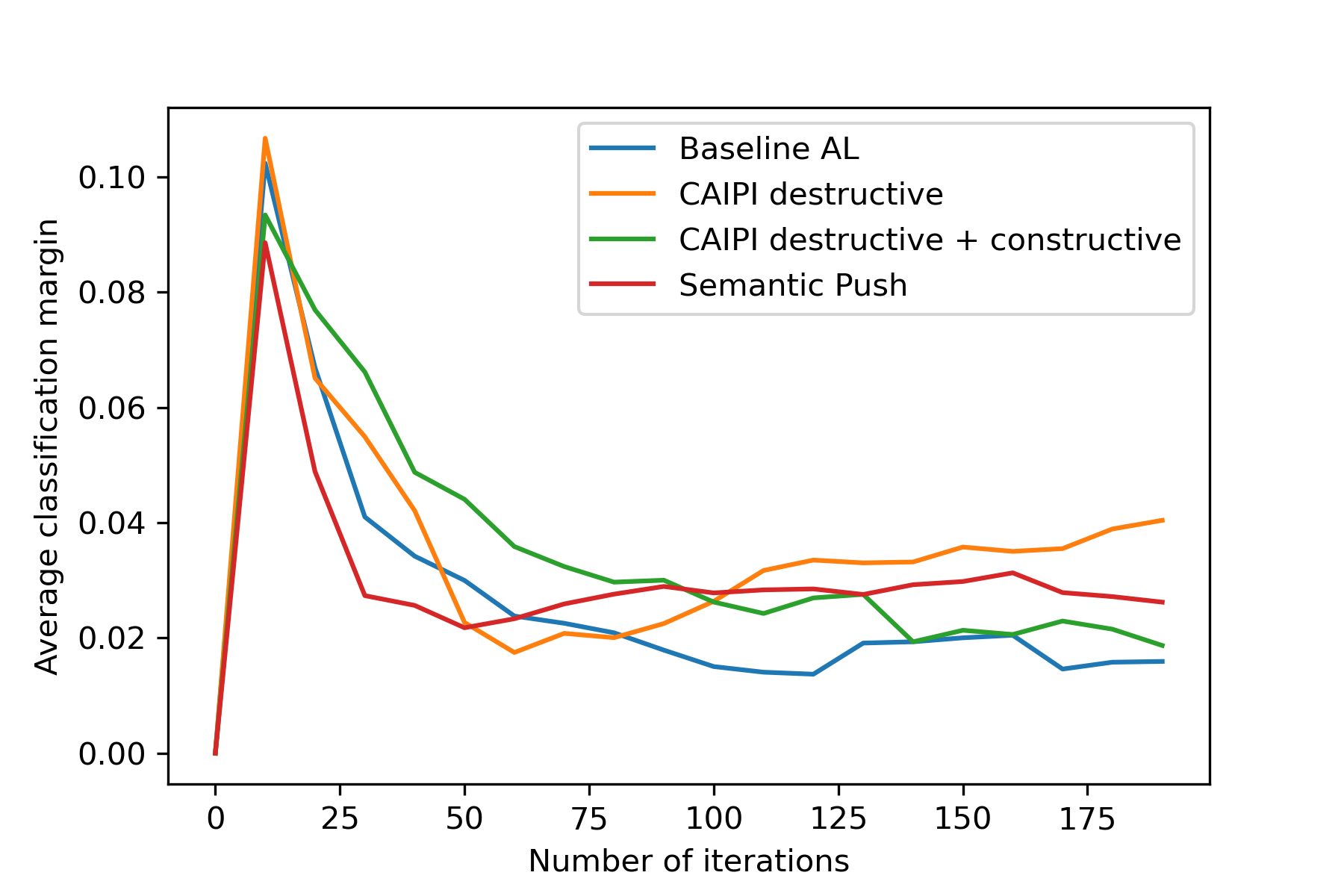}}
	\caption{(a): Learning performance of different IML strategies for Reuters R10 Dataset (Support Vector Machine). (b): Average Classification Margin of different IML strategies for Reuters R10 Dataset (Support Vector Machine).}
	\label{img:PerformanceAndMargin_Reuters_SVC}
\end{figure}

%

\clearpage
\section*{Reproducibility}
To facilitate the reproducibility of this work, our code is available at https://github.com/sb1990gtr/Semantic-Interactive-ML.

\end{document}